\documentclass[conference]{IEEEtran}
\usepackage{blindtext, graphicx, array}
\usepackage{hyperref}
\usepackage{complexity}
\usepackage{fancyhdr}
\usepackage{lastpage}

\ifCLASSINFOpdf
\else
\fi

\usepackage{stfloats}
\usepackage{graphics}
\usepackage{url}
\begin{document}

\title{TaintAssembly: Taint-Based Information Flow Control Tracking for WebAssembly}

\author{\IEEEauthorblockN{William Fu, Raymond Lin, Daniel Inge}
\IEEEauthorblockA{Harvard University \\
\{wfu, rlin, dinge\}@college.harvard.edu}}

\maketitle

\begin{abstract}
 WebAssembly (wasm) has recently emerged as a promisingly portable, size-efficient, fast, and safe binary format for the web. As WebAssembly can interact freely with JavaScript libraries, this gives rise to a potential for undesirable behavior to occur. It is therefore important to be able to detect when this might happen. A way to do this is through taint tracking, where we follow the flow of information by applying taint labels to data. In this paper, we describe TaintAssembly, a taint tracking engine for interpreted WebAssembly, that we have created by modifying the V8 JavaScript engine. We implement basic taint tracking functionality, taint in linear memory, and a probabilistic variant of taint. We then benchmark our TaintAssembly engine by incorporating it into a Chromium build and running it on custom test scripts and various real world WebAssembly applications. We find that our modifications to the V8 engine do not incur significant overhead with respect to vanilla V8's interpreted WebAssembly, making TaintAssembly suitable for development and debugging. 
\end{abstract}

\begin{IEEEkeywords}
WebAssembly, Taint Tracking, V8, Chromium
\end{IEEEkeywords}

\section{Introduction}
There has been a trend toward more demanding applications on the web. This necessitates a delivery format that is portable across platforms, compact in size, fast to execute, and comparatively safe to use. JavaScript is too slow for most complex applications, and it is inconvenient as a compilation target. Various solutions have been proposed for this problem, including \texttt{asm.js}, which has gained reasonable popularity as of late \cite{asmjs}. The newest, cross-browser solution is WebAssembly (wasm) \cite{WASM:overviewpaper}. Although it is executed in a partially restricted environment, wasm can interact freely with JavaScript libraries and functions \cite{WASM:officialwebsite}, providing a vector for wasm to interact with other areas of the users' computer in unpredictable and potentially undesirable ways.

With this in mind, it would be nice for developers to be able to detect possible vulnerabilities and unexpected behavior in their wasm programs. In particular, it would be useful to track data flows between wasm programs and JavaScript, as that is the primary way in which wasm is exposed. This motivates dynamic taint tracking, in which we mark and follow the flow of information through a system.

As far as we know, no tools currently exist for performing such analysis on wasm. Given wasm's current early stage of development, it would be helpful for such a utility to be developed before wasm becomes ubiquitous. We therefore propose TaintAssembly, a modified version of the V8 JavaScript engine that implements dynamic taint tracking for interpreted WebAssembly.

We begin with describing relevant background and previous research on WebAssembly and taint tracking in Section 2. We follow this with a description of the TaintAssembly engine that we implemented in Section 3. Section 4 describes various testing and benchmarking that we performed on our engine, Section 5 described some limitations to our approach, Section 6 denotes some difficulties faced, and Section 7 contains some concluding remarks.

\section{Background and Related Work}
\subsection{WebAssembly (wasm)}

WebAssembly is a new portable, size- and load-time efficient format, suitable for compilation to the web. WebAssembly is not intended to be a standalone programming language, but rather is primarily designed to act as a compile target for C/C++ code (or JavaScript \cite{WASM: crosscompile-javascript}). The wasm specifications \cite{WASM:officialwebsite} lay out the primitives and structures that must be available, and all of the major browsers have implemented some way of executing wasm code. WebAssembly has a ``linear memory," which is analogous to the heap in C/C++. WebAssembly does not enforce memory safety within its memory. However, this memory is kept isolated from everything else, including the code space, execution space, and the executing engine's code and data. WebAssembly also has a standard interpreter, used for testing production code and prototyping new code \cite{WASM:overviewpaper}.

Chromium's V8 JavaScript engine is able to execute wasm, and will be the primary engine considered in this paper. V8 uses two different methods to process wasm bytecode.

\begin{enumerate}
    \item \emph{Interpreted Method:} This is primarily used for debugging and testing, and executes the wasm bytecode line by line at runtime \cite{WASM:overviewpaper}.
    \vspace{2mm}
    \item \emph{Compiled Method:} This compiles a wasm file to native bytecode before execution, mitigating the warm-up time that slows down JavaScript \cite{WASM:overviewpaper}. This means that at execution time, the wasm code runs extremely quickly. This is the default way in which most real-world wasm is handled.

\end{enumerate}

Although V8 compiles most WebAssembly applications to native bytecode, implementing taint tracking for native bytecode is difficult as it would require targeting a user's native architecture (\texttt{mips}, \texttt{arm}, \texttt{x86}, etc.). As this varies from user to user, this would greatly increase the complexity of taint tracking. We foresee our tool as something used primarily by researchers and developers, and not in a production environment, so we will focus solely on interpreted wasm. 

\subsection{Taint Tracking}

Taint tracking is a method for following specific data throughout the execution of a program. Each item that you want to track, be it a primitive or object, is assigned an additional variable that will store its taint value. Each bit of the taint value corresponds to a different source of data, such as network packets, user input, or other information that might be considered important. When these data sources are used in calculations, or interact with each other, a taint tracking engine updates the taint values accordingly, allowing a developer to observe how information has interacted. 

Taint tracking has been implemented previously for a number of different goals and applications. It can be used to detect vulnerabilities in binaries and assist in the analysis of malware and network protocols, as well as actively guard against attacks in runtime \cite{Taint:allyoueverwantedtoknow}. It has been implemented on a variety of systems, ranging from TaintDroid \cite{Taint:taintdroid}, which modified the Dalvik interpreter to perform taint tracking for Java bytecode on Android, to static and dynamic taint analysis implemented for JavaScript bytecode in Safari's Webkit Javascript engine \cite{Taint:javascript}. 

Taint tracking has also been implemented at a lower level. Panorama \cite{Taint:panorama} implements taint tracking at the instruction level for a whole system with fine granularity using QEMU. They use shadow memory, to store taints for every byte of memory, all registers and the network buffer. They also implement conditional taint propagation for certain functions, notably those taking keyboard input.

A lot of work has been done to speed up taint tracking for both debugging and production use. Some approaches to taint tracking use heuristics to determine the granularity of dynamic taint tracking. For example, the CloudTaint system \cite{Taint:cloud} uses certain triggers from data sources to decide whether or not to activate taint tracking at the instruction level. This prevents the CloudTaint system from constantly using up processing power needed for other VMs in the group. 

Unfortunately, static taint analysis suffers somewhat from a lack of precision while dynamic taint tracking has a high performance overhead \cite{Taint:fastdynamictaint}. Therefore, most recent work into taint tracking has taken a combined approach. 

One such ensemble approach is to pre-process taint tracking for specific functions, establishing function summaries. Zhu et. al. \cite{Taint:tainteraser} use the pre-computed function summaries along with semantic analysis \cite{IFC:languagebased} \cite{Taint:straighttaint} to speed up taint tracking by an order of magnitude. Through static analysis, some functions can be determined to not propagate taints from certain inputs to outputs. Instead of running through the entire function with full taint propagation, only a patch function is needed to propagate taint from the inputs to the outputs, eliminating most of the overhead and context switching associated with propagating taint. While most of the previous work has dealt with compiled binaries, the idea of using function summaries to speed up dynamic taint propagation is still applicable to wasm binaries.

Another approach that has worked well is not optimization through static analysis of binaries or code, but rather a dynamic optimization of taint propagation itself. Bruening et al. \cite{Taint:dynamicoptimization} constructed a dynamic optimization system that optimizes linear sequences of code with changing levels of details to describe instructions. Although Bruening's work did not directly involve taint propagation, similar ideas can be applied to dynamic taint tracking as well. Saxena et al. \cite{Taint:binaryinstrumentation} demonstrated a method of storing a metadata stack and implemented metadata caching in registers, which allowed for quick retrieval of small metadata such as taint during runtime. 

In our TaintAssembly engine, our basic taint tracking is modeled off of TaintDroid \cite{Taint:taintdroid}, while our taint tracking on the wasm linear memory draws from the idea of shadow memory presented in Panorama \cite{Taint:panorama}. The other work presents interesting ideas for optimizing and improving dynamic taint tracking, but are significantly trickier to implement in the WebAssembly interpreter, so for the sake of time and simplicity we do not implement these in our TaintAssembly engine.

\section{Implementation}
We implemented taint tracking for WebAssembly by modifying the V8 JavaScript Engine, used in Google Chrome, Chromium, and Node.js  \cite{WASM:v8website}. A GitHub repository containing our modifications, as well as some testing scripts, can be found at \url{https://github.com/wfus/WebAssembly-Taint}.

\subsection{Basic Taint Tracking}

Our first modification was implementing basic taint tracking. For interpreted WebAssembly, V8 wraps the WebAssembly standard values \texttt{i32}, \texttt{i64}, \texttt{f32}, \texttt{f64} using a \texttt{WasmValue} class. Therefore, for caching purposes, we introduced taint labels as part of each \texttt{WasmValue} object, similar to TaintDroid's modification to Dalvik \cite{Taint:taintdroid}. Our structure is exhibited in Figure 1. Note that while we allow the user to set the size of \texttt{taint\_t}, our default is $4$ bytes, which is what we have exhibited in the diagram, and what we will assume for the rest of this paper.

\begin{figure}[h]
    \centering
    \includegraphics[scale=0.6]{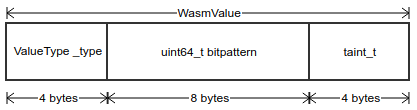}
    \caption{The structure of a \texttt{WasmValue} with taint appended.}
    \label{fig:wasmvalue}
\end{figure}

In order to initialize the taints, we implemented signature overloading, so a user simply needs to add extra parameters describing taint values to their function calls. For example, if we have a sample JavaScript function invoking a wasm function
\begin{verbatim}
myfunction = exports._wasm_function 
\end{verbatim} 
with three integer parameters $A$, $B$, $C$, we could have function calls
\begin{center}
\begin{verbatim}
  myfunction(50, 100, 200);
  myfunction(50,100, 200, 0x000000f0);
  myfunction(1, 2, 3, 0x1, 0x2, 0x4, 0x8);
\end{verbatim}
\end{center}
where in the first line, all parameters have no taint, in the second the first parameter has taint \texttt{0x000000f0}, and in the third the parameters have taint \texttt{0x1}, \texttt{0x2}, and \texttt{0x4}, respectively. Note that in the third example, the \texttt{0x8} is thrown away since there are more taint labels provided than there are parameters. 
\begin{table*}[t]
\begin{center} 
\vspace{1mm}
\resizebox{1\textwidth}{!}{
\begin{tabular}{l l l l}
\hline
    \textbf{Wasm Opcode} 
    & \textbf{Operation} 
    & \textbf{Taint Propagation} 
    & \textbf{Description} \\
\hline
    0x45 - 0x4f  & i32.binop $v_1$ $v_2$ (comparison) 
        & $T(R) = 0$ & Comparison ops do not propagate taint.\\
    0x50 - 0x5a  & i64.binop $v_1$ $v_2$ (comparison) 
        & $T(R) = 0$ & Comparison ops do not propagate taint.\\
    0x5b - 0x60  & f32.binop $v_1$ $v_2$ (comparison) 
        & $T(R) = 0$ & Comparison ops do not propagate taint.\\
    0x61 - 0x66  & f64.binop $v_1$ $v_2$ (comparison) 
        & $T(R) = 0$ & Comparison ops do not propagate taint.\\
    0x67 - 0x69  & i32.unop $v_1$ 
        & $T(R) = T(v_1)$ & Unops should receive taint of the parameter.\\
    0x70 - 0x78  & i32.binop $v_1$ $v_2$ (non-comparison)
        & $T(R) = T(v_1) \lor T(v_2)$ & Non-comparison binops should receive taint of both parameters. \\
    0x79 - 0x7b & i64.unop $v_1$ 
        & $T(R) = T(v_1)$ & Unops should receive taint of the parameter.\\
    0x7c - 0x8a  & i64.binop $v_1$ $v_2$ (non-comparison)
        & $T(R) = T(v_1) \lor T(v_2)$ & Non-comparison binops should receive taint of both parameters. \\
    0x8b - 0x91  & f32.unop $v_1$ 
        & $T(R) = T(v_1)$ & Unops should receive taint of the parameter.\\
    0x92 - 0x98  & f32.binop $v_1$ $v_2$ (non-comparison)
        & $T(R) = T(v_1) \lor T(v_2)$ & Non-comparison binops should receive taint of both parameters. \\
    0x99 - 0x9f  & f64.unop $v_1$ 
        & $T(R) = T(v_1)$ & Unops should receive taint of the parameter.\\
    0xa0 - 0xa6  & f64.binop $v_1$ $v_2$ (non-comparison)
        & $T(R) = T(v_1) \lor T(v_2)$ & Non-comparison binops should receive taint of both parameters. \\
\hline
\end{tabular}} 
\end{center}
\caption{Taint Propagation Logic}
\end{table*}
\renewcommand{\arraystretch}{1}

The reason for overloading function signatures as a method of inputting taint is to allow for compatibility between TaintAssembly and the clean default V8 engine. Clean implementations of V8 will ignore overloaded arguments, without error. Therefore, users of TaintAssembly are able to modify any wasm functions in their code to inject taint, while still being able to run the original, unchanged files in clean V8.

Finally, we made modifications to the wasm interpreter to implement the taint semantics similar to the ones described in other taint tracking engines. We changed the way that the interpreter handles binary and unary operations to include taint propagation. Specifically, as exhibited in Table 1 (where $R$ denotes the result of the operation), for all non-comparison operators, we want to generate the taint of the output using the taints of the inputs. This means that for non-comparison binary operations, the taint of the output is the bitwise OR of the taints of the inputs. For unary operations, the taint of the output is simply the taint of the input. With comparison operators, we do not propagate taint.

\subsection{Tainting Linear Memory}

WebAssembly utilizes linear memory, analogous to the heap in C, with the operations \texttt{T.load} with \texttt{T} being one of the four primitives. In V8, both the wasm interpreter and compiler need to be able to write to and read from arbitrary memory addresses in linear memory, which makes it difficult to store the taint alongside data. For example, we have here a simple C program could be converted into the corresponding \texttt{.wasm} binary using the WebAssembly Explorer \cite{benchmark:wasmexplorer}: 
\begin{verbatim}
void rudewrite(int* r, int N) {
    for (int i = 0; i < N; i++) {
        memcpy(r, i, sizeof(int));
        r++; 
    }
}
\end{verbatim}
\begin{verbatim}
(i32.load align=1
    (i32.add
        (get_local $2)
        (i32.const 8)))
\end{verbatim}
This computes a memory address by adding a parameter and a constant value, and loads a value located in linear memory at that address. Since it offsets by \texttt{sizeof(int)}, it would be difficult to place taints alongside data without substantial revision to the original code. Using a function to map memory locations with spacing left for taint would also be difficult, because our \texttt{taint\_t} type defaults to 4 bytes while single byte values are common in C. 

Therefore, similarly to Panorama's shadow heap \cite{Taint:panorama}, we implemented a mapping from memory addresses to taint values. Instead of using page tables, we implemented it using an \texttt{std::unordered\_map}. Whenever a memory value is read in through \texttt{T.load}, the address is checked to see if it has taint before wrapping it inside a \texttt{WasmValue}. Likewise, when a \texttt{T.store} command is executed, the WasmValue's taint is unwrapped and the address and taint placed as key and value in our map. Whenever a wasm context destroys itself, it clears the map. This design is illustrated in Figure 2.

\begin{figure}
    \centering
    \includegraphics[scale=0.7]{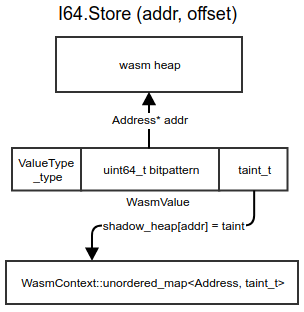}
    \caption{Tracking taint inside the wasm linear memory (heap). Note the unique WasmContext }
    \label{fig:my_label}
\end{figure}

We originally meant to modify V8's memory allocation for linear memory be allocating a section of memory opposite of linear memory for storing taints. We decided to use a C++ \texttt{std::unordered\_map}, however, because V8 uses a generic Handler for managing memory allocations and offsets. Given the time constraints, we were unable to ensure that our tainted linear memory would grow with the regular linear memory without trampling over other V8 internal memory structures. Therefore, our current implementation uses a \texttt{unordered\_map} kept for each \texttt{WasmContext}.

\subsection{Probabilistic Taint Tracking}

Finally, we included an option for probabilistic taint, where taint propagates only some of the time. The propagation semantics are as follows. Let us have values $v_1,v_2$ with respective taints $t_1,t_2$ and probabilities of propagation $p_1,p_2$ and operation $op$. Let $r$ be the result of operation $op$ on those values. Then, $r$ will have taint $t_1 \lor t_2$ with probability $p_1p_2$, $t_1$ with probability $p_1(1-p_2)$, $t_2$ with probability $(1-p_1)p_2$, and $0$ with probability $(1-p_1)(1-p_2)$. In other words, the taint from value $v_i$ is propagated with probability $p_i$. Finally, the propagation probability associated with this $r$ will be $\max(p_1,p_2)$. 

In terms of representing the taint value and probability of propagation, we preserve the same taint label structure as from basic taint tracking. The difference is that we reserve some number of higher order bits of the taint label for representing the propagation probability, as in Figure 3.

\begin{figure}[h]
    \centering
    \includegraphics[scale=0.6]{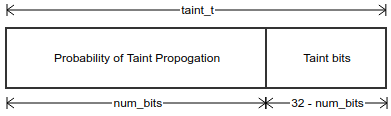}
    \caption{The structure of a \texttt{taint\_t} when probablistic taint propagation is being used.}
    \label{fig:taint}
\end{figure}

The way this encoding works is that if there are $n$ bits reserved for the probability and the value stored there is $m$, the propagation probability being represented is $p = \frac{m}{2^n - 1}$.

The motivation for probabilistic taint propagation is primarily that it allows us to set some sort of ``lifetime" for taint. As a taint gets passed through the data, the propagation probability makes it possible for the taint to disappear over time. Specifically, the smaller the propagation probability, the shorter a taint's lifetime is. This might desirable if you have some value with unimportant taint, and you only wish for it to only taint values that it acts upon most directly.

\subsection{Additional Features}

In addition to basic taint tracking, we also implemented logging. We provide two levels of detail to our logging. In one, we only log when a tainted value is returned from WebAssembly to JavaScript, while in the other we log all operations and function calls.

We also wrote functionality that would terminate execution of a program if a tainted value is returned from WebAssembly to JavaScript. Specifically, we provided an option for terminating the program on the appearance of return values with specific taint flags set, where the relevant flags would be user-input. For instance, a user would input \texttt{0x15} if they would like we could termination whenever an output with taint \texttt{0x1}, \texttt{0x4}, or \texttt{0x10} is returned.

The primary motivation for this feature is giving the developer a nice and easy way of seeing when tainted values are getting inadvertently passed to JavaScript libraries without having to go through a taint log. In addition, although our taint tracking engine is primarily meant for developer and debugging use, this feature could hypothetically allow a user to prevent their own information from being leaked out by running our taint tracking engine.



\section{Performance Evaluation}

\begin{table*}[t]
\begin{center}
\addtolength{\tabcolsep}{-2pt}   
\resizebox{1\textwidth}{!}{
    \begin{tabular}{|m{3cm}|p{9cm}| }
    \hline
         \textbf{Program} & \textbf{Description} \\
    \hline
         Factorial \cite{benchmark:factorial} & 
            A basic page that advertises the new Rust-To-Wasm (TM) compiler! As everyone knows, if something already exists, let's rewrite it in rust! In any case, this is just a simple web page that uses wasm to calculate the factorial function.
            \\\hline
         Skeletal Simulation \cite{benchmark:skeletal} &  
            A WebGL simulation of spooky dancing humanoid figures that compares rendering with WebAssembly and Javascript.
            \\\hline
         Video Editor  \cite{benchmark:videoeditor}&    
            Basic video editor demo that takes in webcam input. Applies zany filters to the video and compares the time it takes for WebAssembly and JavaScript to perform the same filters. 
            \\\hline
        Funky Karts \cite{benchmark:funkykarts} & 
            A WebGL game that utilizes WebAssembly. It's a fun side-scrolling racing game featuring a badger driving through treacherous terrain on in a wild race to save his missing friends. Uses cartoon type graphics that does not seem resource intensive. 
            \\\hline
        
    \end{tabular}} 
\end{center}
\caption{Testing Programs}
\end{table*}

In this section, we describe various tests and benchmarks that we used to evaluate our TaintAssembly engine. We performed tests involving a variety of web applications employing WebAssembly to assess the runtime overhead resulting from our modifications, as well as an examination of a specific source of overhead arising from our method for introducing taint. Finally, we do some analysis of our probabilistic taint feature, looking at how propagation probability relates to taint lifetime.

\subsection{Program Runtime}
The basic taint tracking features of our TaintAssembly engine were tested on some online applications that use WebAssembly. Since TaintAssembly is essentially a modified version of V8's \texttt{chromium/3270} branch, all testing was performed on a Chromium build (version \texttt{64.0.3270.2}) on Ubuntu 17.1 with the custom V8 engine. As a comparison, we used a clean build of Chromium version \texttt{64.0.3270.2} to benchmark against. 

In our analysis, we noted down the runtime of the following
\begin{enumerate} 
    \item Compile wasm on clean V8 engine
    \item Interpreted wasm on clean V8 engine
    \item Interpreted wasm on TaintAssembly engine (no taint passed in)
    \item Interpreted wasm on TaintAssembly engine (taint passed in)
\end{enumerate}

The specific applications that we used for benchmarking are described in Table 2. 

The definitions that we used for runtime in each of the applications are as follows. For the factorial, we defined runtime to be the amount of time that it took for the wasm factorial function to compute a particular factorial. For the skeletal simulation, we let the runtime be the amount of time that it took to perform the calculation for a step of the animation. For the video editor, we chose to analyze one of its filters, the invert filter. We set runtime here to be the time it took to call the wasm function performing the calculations for the transformation on each pixel in the frame. For Funky Karts, we used the in-game clock to help us measure runtime. In particular, we timed the amount of real time that corresponded with three seconds in game time.

The results from our tests are presented in Tables 3 and 4. In Table 3, we present the runtimes relative to the clean V8 engine with compiled wasm, and in Table 4, we present them relative to the clean V8 engine with interpreted wasm. We notice that the factorial and Funky Karts programs have reasonable overhead relative to compiled wasm on a clean V8 engine, while the overheads for the Skeletal Simulation and Video Editor are substantial. However, from looking at the runtimes relative to interpreted wasm on a clean V8, the overheads for TaintAssembly are all fairly reasonable. We observe that much of the apparent slowness of TaintAssembly versus compiled WebAssembly arises from simply using interpreted WebAssembly, and that our taint tracking modifications do not add large overhead (generally only around $5-12\%$) to the engine. Therefore, it is reasonable for someone to use our TaintAssembly engine instead of a vanilla wasm interpreter when developing.

\begin{table*}[t]
\begin{center}
\resizebox{1\textwidth}{!}{
\begin{tabular}{|c ||c |c |c |c| c|c |}
\hline
    \textbf{Program} & \textbf{Clean (compiled)} & \textbf{Clean (interpreted)}  &\textbf{TaintAssembly (untainted)} & \textbf{TaintAssembly (tainted)} \\\hline
    Factorial  
        &   1.000
        &   2.381
        &   2.400
        &   3.562
        \\\hline
    Skeletal Simulation
        &   1.000
        &   230.644
        &   247.088
        &   240.960
        \\\hline
    Video Editor
        &  1.000
        &  200.342	
        &  208.261	
        &  216.702
        \\\hline
    Funky Karts 
        & 1.000
        & 9.695	
        & 10.835	 
        & 10.839
        \\\hline
\end{tabular}}
\end{center}
\vspace{2mm}
\caption{Benchmarking Results (relative to Clean (compiled))}
\end{table*}
\begin{table*}[t]
\begin{center}
\resizebox{0.8\textwidth}{!}{
\begin{tabular}{|c ||c |c |c |c|}
\hline
    \textbf{Program} & \textbf{Clean (interpreted)}  &\textbf{TaintAssembly (untainted)} & \textbf{TaintAssembly (tainted)} \\\hline
    Factorial 
    &   
        1.000 &  
        1.008 & 1.496
         \\\hline
    Skeletal Simulation &  
        1.000 &  
       1.071 &1.045
    \\\hline
    Video Editor &  
        1.000 &
        1.040& 1.082
     \\\hline
    Funky Karts & 
        1.000 &
        1.118& 1.118 \\\hline
\end{tabular}}
\end{center}
\vspace{2mm}
\caption{Benchmarking Results (relative to Clean (interpreted))}
\end{table*}

\subsection{Taint Insertion Overhead}

One of the primary sources of overhead is the introduction of taint through parameter overloading, so we would like to see exactly how much slowdown occurs. To do this, we wrote simple C scripts that did not do any operations and simply took in parameters. We then converted these to wasm with the WebAssembly Explorer \cite{benchmark:wasmexplorer}. Then, we timed executions of functions calls to the resultant wasm binary on both a clean, unmodified V8 engine and our TaintAssembly engine. More specifically, the functions that we tested each took in $100$ arguments. We had one function taking in all \texttt{INT32}, one taking in all \texttt{FLOAT32}, and one taking in all \texttt{FLOAT64}. Note that we do not have any results for \texttt{INT64} since $64$ bit integer types are not currently supported in JavaScript \cite{WASM:overviewpaper}. The results are in Figure 4.

\begin{figure}[h]
    \begin{center}
    \includegraphics[scale=0.4]{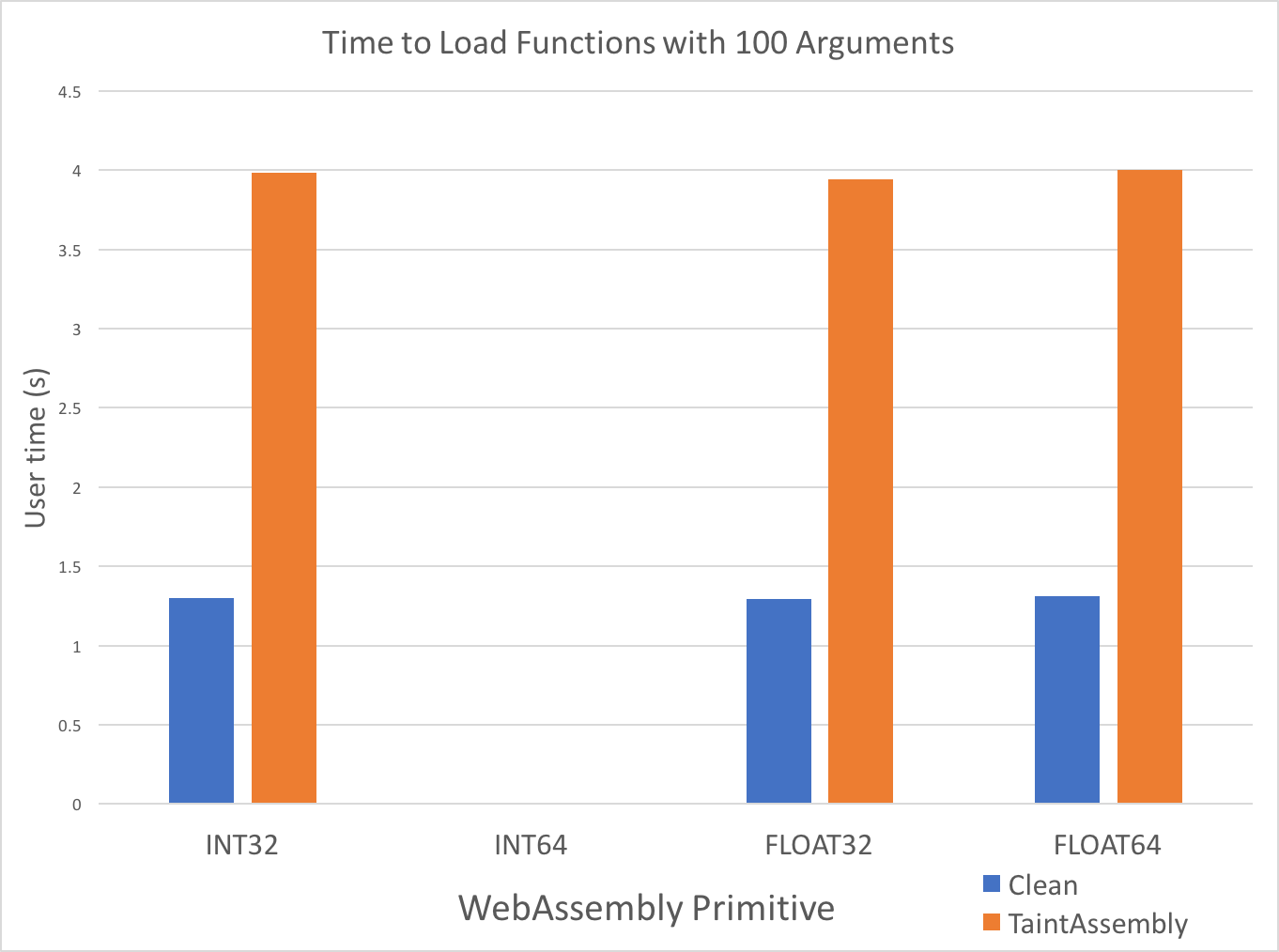}
    \caption{Overhead of passing taint parameters to functions with 100 arguments.}
    \label{fig:100args}
    \end{center}
\end{figure}
In an extreme case with 100 parameters, we note that our taint tracking engine is only around $3$ times slower than the clean engine. Although there is overhead from injecting taint into our function, the overhead is small enough for development and debugging purposes. Most of the additional time comes from using the function signature to pass in the taint; it allows for minimal modifications to the source code when passing in taint, with the cost of some small overhead. 

\subsection{Probabilistic Taint Benchmarks}

We analyze our probabilistic taint feature. As a test of this propagation, we pass tainted data through a simple hash function from \cite{benchmark:hash}, which takes a \texttt{uint32\_t} and returns a \texttt{uint32\_t}. The hash function does a total of $2700$ binary operations. We implemented the hash function in C and converted to wasm with the WebAssembly Explorer \cite{benchmark:wasmexplorer}. Using a resolution of $8$ bits for the probability and running each possible probability $k$ for $100000$ iterations, we can see the approximate lifetime of our taint given a specific probability. To quantify this lifetime for a given propagation, we use the probability that applying the hash function to tainted values results in a tainted output. The resultant plot is exhibited in Figure 5. 

\begin{figure}[h]
    \centering
    \includegraphics[scale=0.5]{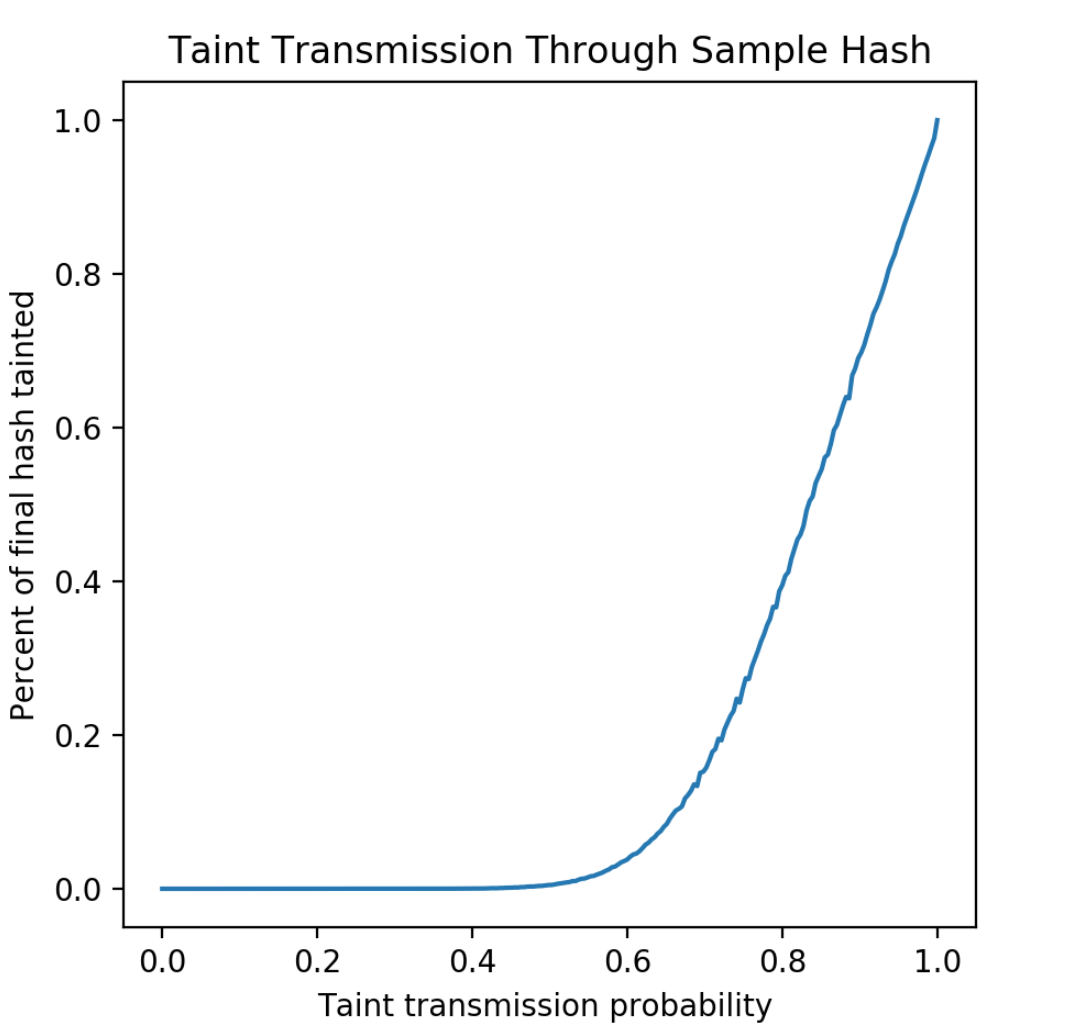}
    \caption{Percent of return values that retain taint after passing through a simple hash function}
    \label{fig:hashplot}
\end{figure}

In general, the exact shape of this plot is determined by the exact hash function and the number of passes used. In this particular case, it appears to approximate a softplus function. We see that there is a quick drop-off in taint lifetime if the propagation probability goes below $0.6$, and that we have around a $50\%$ chance of seeing a tainted result if the propagation probability is around $0.8$.

\section{Limitations}

\subsection{Approach Limitations}

Our taint propagation semantics explicitly do not assign taint to comparison operators, as the simplest forms of such taint transmission readily result in taint explosion. However, this has the side effect of not propagating indirect taint. As in Panorama \cite{Taint:panorama}, we could have handled this by only tainting the program counter to propagate indirect taint in specific functions, but we did not have a chance to implement such a solution due to time constraints.

\subsection{Taint Source Limitations}

In our implementation, our option for probabilistic taint takes up some of the most significant bits of our taint, and requires random number generation. This has some potential drawbacks:
\begin{enumerate}
    \item Probability resolution is limited by the number of bits 
    \item Some taint resolution is sacrificed for taint lifetime information
    \item A \texttt{rand()}\footnote{V8's default settings are not optimal as they require the user to seed their own entropy source.} call and \texttt{min} calculation are required per-op
\end{enumerate}
Therefore, for some computationally heavy programs with lots of taint sources and binary operations, it may not be feasible to run TaintAssembly with probabilistic taint.

\subsection{WebAssembly Linear Memory}

Unfortunately, it was difficult to implement efficient taint tracking for WebAssembly's linear memory. Since memory locations can be manipulated directly, for complete taint tracking we need to create another data structure to taint values in linear memory \cite{Taint:panorama}. The current implementation causes a slight runtime overhead for C/C++ programs that frequently read from and write to the heap and has a large memory overhead.

\section{Challenges}
Much of our time was spent looking at the V8 source code. Since the code for the runtime was complex and linked with most of the other source files, the builds took a long time to compile. Furthermore, we had to trace through the source code manually and test with the debug shell for most of our initial testing. Tracing the program calls of the default wasm interpreter was also difficult, because V8's JIT caused backtraces to fail.  

We originally attempted to follow V8's master branch, but due to active development in \texttt{master} we had to rewrite our code to target a stable branch. We ultimately ended up choosing the V8 branch \texttt{chromium/3270} and Chromium version \texttt{64.0.3270.2} \cite{WASM:chromiumwebsite}.

\section{Conclusion}

In this paper, we have implemented dynamic taint tracking for interpreted WebAssembly in the V8 JavaScript Engine. We modified the V8 wasm interpreter, allowing it to perform basic taint tracking with function parameters and local variables, taint tracking in linear memory, and probabilistic taint tracking. In addition, we tested our modified engine on some web applications using wasm and our own custom scripts. Compared to the default interpreter, our TaintAssembly engine has a reasonable overhead of around $5-12\%$ for all of the web applications we tested. Therefore, we have exhibited a simple WebAssembly taint tracking engine that is suitable for development and debugging.


%

\section*{Acknowledgments}

The authors would like to thank James Mickens for his guidance and his excellent lectures on systems security. The authors would also like to thank the anonymous members on the Google V8-users and V8-devs mailing list for standing in for V8's documentation, for no benefit of their own. Without their help, we would never have been able to navigate through V8's internal code structure without significant difficulty.

\ifCLASSOPTIONcaptionsoff
  \newpage
\fi



%

%

\begin{IEEEbiography}[{\includegraphics[width=1in,height=1.25in,clip,keepaspectratio]{picture}}]{John Doe}
\blindtext
\end{IEEEbiography}




\end{document}